\documentclass[prl,amsmath,aps,twocolumn,superscriptaddress]{revtex4}
\usepackage{amsfonts}
\usepackage{graphicx,multirow,color}

\begin{document}
\title{All entangled pure states violate a single Bell's inequality}
\author{Sixia Yu}
\affiliation{Centre for quantum technologies, National University of Singapore, 3 Science Drive 2, Singapore 117543, Singapore}
\affiliation{
Hefei National Laboratory for Physical Sciences at Microscale and Department of Modern Physics,  University of Science and Technology of China, Hefei, Anhui 230026, China}
\author{Qing Chen}
\affiliation{Centre for quantum technologies, National University of Singapore, 3 Science Drive 2, Singapore 117543, Singapore}
\author{ Chengjie Zhang}
\affiliation{Centre for quantum technologies, National University of Singapore, 3 Science Drive 2, Singapore 117543, Singapore}
\author{C.H. Lai}
\affiliation{Centre for quantum technologies, National University of Singapore, 3 Science Drive 2, Singapore 117543, Singapore}
\affiliation{Physics department, National University of Singapore, 3 Science Drive 2, Singapore 117543, Singapore}
\author{C.H. Oh}
\affiliation{Centre for quantum technologies, National University of Singapore, 3 Science Drive 2, Singapore 117543, Singapore}
\affiliation{Physics department, National University of Singapore, 3 Science Drive 2, Singapore 117543, Singapore}

%\author{Sixia Yu, Qing Chen, Chengjie Zhang, C.H. Lai, and C.H. Oh}
\begin{abstract}
We  show that a single Bell's inequality with two dichotomic observables for each observer, which originates from Hardy's nonlocality proof without inequalities, is violated by all entangled pure states of a given number of particles,  each of which may have a different number of energy levels. Thus Gisin's theorem is proved in its most general form from which it follows that for pure states Bell's nonlocality and quantum entanglement are equivalent.
\end{abstract}

%\pacs{03.65.Ta, 03.67.-a}
\maketitle

Quantum nonlocality as revealed by the violations of various Bell's inequalities \cite{bell} is intriguingly related to quantum entanglement. On the one hand Werner \cite{werner} showed that there exist entangled {(mixed)} states that can be simulated by local hidden variable models and thus cannot exhibit any nonlocality in the manner of Bell. On the other hand  Gisin  \cite{gisin}  showed that all the entangled pure states of two qubits violate a single Bell's inequality, namely the Clause-Horne-Shimony-Holt (CHSH) inequality \cite{CHSH}, with two different measurement settings for each observer. This result is referred to as Gisin's theorem and ever since there have been many efforts \cite{pr,zb,chen,3,gisinperes,chen2,shao,commt}, successful and unsuccessful, to generalize Gisin's theorem to multipartite systems with multilevels, trying to establish an equivalency between the quantum entanglement and Bell's nonlocality for pure states.

The first effort to generalize Gisin's theorem to multipartite systems was made by Popescu and Rohrlich~\cite{pr} who showed that all the entangled pure multipartite states violate a set of conditional Bell inequalities. Since postselections are involved,  {as noticed by \.{Z}ukowski {\it  et al.} \cite{zb},} their approach cannot be regarded as a valid proof of Gisin's theorem for multipartite system. Also  it was shown that Bell inequalities for {{\it full correlations}} with two dichotomic observables for each observer cannot reveal the nonlocality of all entangled pure $n$-qubit states \cite{zb}. A breakthrough was made by Chen {\it et al.} \cite{chen} who showed numerically that all 3-qubit pure entangled states violate a Bell inequality for {{\it probabilities}} with an analytical proof given by Choudhary {\it et al.} \cite{3}, showing that a single Bell's inequality with two nondegenerate measurement settings is violated by all entangled pure states of three qubits. As to higher dimensional systems, Gisin and Peres proved \cite{gisinperes} that all the entangled pure  states of two qudits also violate the CHSH inequality and an alternative proof is given by Chen {\it et al.} \cite{chen2}.

Recently a tentative proof of Gisin's theorem for multiparticles with arbitrary number of energy levels was given by Li and Shao \cite{shao}. They showed that for every entangled pure state of multiparticles there exists a partition of particles into two groups such that a bipartite Bell'{s} inequality with two trichotomic observables for each group is violated.  Unfortunately, to obtain violations to their inequalities one needs to measure some collective observables for each effective party which may involve several particles. This  violates the multipartite locality as commented on by Choudhary {\it et al.} \cite{commt}. Thus a proof for Gisin's theorem in general is still missing.

In this Letter we shall prove Gisin's theorem in its most general form by showing that all the entangled pure states violate a single Bell'{s} inequality with two different measurement settings for each observer. After a short introduction to this special Bell's inequality, referred to as Hardy's inequality since it originated from Hardy's proof of nonlocality without inequality, we shall at first demonstrate its violation by an arbitrary entangled pure state of $n$ qubits. We then reduce the problem of finding its violation by an entangled pure state of multiparticles, each of which may have a different number of energy levels, to that of an effective set of $n$ qubits obtained by locally projecting $n$ qudits to $n$ qubits.

{\it Hardy's inequality --- } Consider a system composed of $n$ spacelike separated subsystems that are labeled with the index set $I=\{1,2,\ldots,n\}$. In any  local realistic model the value of an observable of any subsystem is determined by some hidden variables $\lambda$, distributed according to $\varrho_\lambda$, and is independent of which observables might be measured on other subsystems as required by multipartite locality. For each subsystem ${k\in I}$ we choose two observables $\{a_k,b_k\}$ taking binary values $\{0,1\}$, and Hardy's inequality reads
\begin{eqnarray}\label{d22}
\langle H\rangle_c:=\int d\lambda\varrho_\lambda H\le 0,\quad H=a_I-\bar b_I-\sum_{k\in I} b_ka_{\bar k},
%\le a_I\left(1-\sum_{k\in I} b_k-\bar b_I\right)\le 0.
\end{eqnarray}
where we have denoted $a_\alpha=\prod_{k\in \alpha}a_k$ and $\bar b_\alpha=\prod_{k\in \alpha} \bar b_k$ with $\bar b_k=1-b_k$  for an arbitrary subset $\alpha\subseteq I$ and  $\bar k=I\setminus\{k\}$ for arbitrary $k\in I$.
Based on Hardy's proof of nonlocality without inequality \cite{hardy} Mermin  formulated Hardy's inequality for two qubits \cite{mermin} which was generalized to $n$ qubits by Cereceda \cite{hardyn}. Hardy's inequality is a Bell's inequality for {\it probabilities} and, as it stands, is applicable for a system of $n$ particles each of which may have a different number of energy levels.

Hardy's nonlocality proof can be regarded as a state-dependent proof of quantum contextuality  for $n$ qubits using $2n$ rays. To see this let us consider any noncontextual value assignment to $2n$ binary observables $\{a_k,b_k\}_{k\in I}$ in the spirit of Kochen and Specker \cite{ks} and it is impossible to have  $\bar b_I=b_k a_{\bar k}=0$ for all ${k\in I}$ while $a_I=1$ if the product rule of value assignments holds.
This is because $a_I=1$ leads to $a_k=1$ for all $k\in I$ and from $b_k a_{\bar k}=0$ it follows that $b_k=0$ for all $k\in I$ and thus $\bar b_I=1$, a contradiction. As a result in any noncontextual theory $H$ can never attain a positive value so that Hardy's inequality holds.

For a given entangled pure state $|\psi\rangle$ of $n$ particles, also labeled with $I$, to violate Hardy's inequality Eq. (\ref{d22}) one must find out two measurement settings $\{|a_k\rangle,|b_k\rangle\}$ for each particle $k\in I$ such that
\begin{equation}\label{qd}
\langle H\rangle_\psi:=|\langle \psi|a_I\rangle|^2-|\langle\psi |\bar b_I\rangle|^2-\sum_{k\in I}|\langle \psi|a_{\bar k} b_k\rangle|^2>0,
\end{equation}
where $|a_I\rangle=\otimes_{k\in I}|a_k\rangle_k$, $|\bar b_I\rangle=\otimes_{k\in I}|\bar b_k\rangle_k$ with $|\bar b_k\rangle_k$ being  orthogonal to $|b_k\rangle_k$, and
$|a_{\bar k}b_k\rangle=\otimes_{i\not=k}|a_i\rangle_i\otimes|b_k\rangle_k$. Hardy's  nonlocality proof, in which the measurement settings are so chosen that only the first term of $\langle H\rangle_\psi$ is nonvanishing, provides a natural violation to Hardy's inequality. However not all entangled pure states, e.g., maximally entangled bipartite states \cite{hardy} and a subset of 3-qubit states \cite{wu}, can have  Hardy's nonlocality proof. On the other hand  Hardy's inequality, being  equivalent to the CHSH inequality in the case of two particles, is violated by all the entangled pure bipartite states \cite{gisin,gisinperes}. The analytical proof of Gisin's theorem for 3 qubits \cite{3} is also based on Hardy's inequality, which is found to be violated by all the entangled symmetric pure states of $n$ qubits \cite{sym}.  Here we shall demonstrate that Hardy's inequality is violated by all entangled pure states.

{\it Violations for qubits.--- } We consider at first $n$ qubits, labeled with the index set $I$, and take an arbitrary basis $\{|0\rangle_k,|1\rangle_k\}$ for each qubit $k\in I$ so that $\{|0_\alpha 1_{\bar \alpha}\rangle\}_{\alpha\subseteq I}$ form a basis for $n$ qubits where $|0_\alpha\rangle=\otimes_{k\in\alpha}|0\rangle_k$ and $|1_{\bar \alpha}\rangle=\otimes_{k\in\bar\alpha}|1\rangle_k$ with $\bar\alpha=I\setminus\alpha$ for an arbitrary $\alpha\subseteq I$.
A basis, in which  a given state $|\psi\rangle$ is expanded as
\begin{equation}\label{ex}
|\psi\rangle=\sum_{\alpha\subseteq I}h^*_\alpha|0_\alpha1_{\bar\alpha}\rangle, \quad h_\alpha=\langle\psi|0_\alpha1_{\bar\alpha} \rangle,
\end{equation}
is called a {\it magic basis} for $|\psi\rangle$ if $h_I\not=0$ and $h_{\bar k}=0$ for all $k\in I$ with $\bar k=I\setminus\{k\}$.
By a suitable choice of the local basis for each qubit, a magic basis can always be found.
For example we can construct a magic basis for a given pure state $|\psi\rangle$ with the help of its {\it closest product state} $|p_I\rangle=\otimes_{k\in I}|p_k\rangle_k$ whose inner product with $|\psi\rangle$ is the largest among all possible product states. The closest product state always exists, albeit difficult to find, and makes the definition of the geometric measure of entanglement \cite{gm} possible. Let $|\bar p_k\rangle_k$ be the state orthogonal to $|p_k\rangle_k$  for each qubit $k\in I$, then $\{|p_\alpha \bar p_{\bar \alpha}\rangle\}_{\alpha\subseteq I}$ is a magic basis for $|\psi\rangle$. This is because $|h_I|^2>0$ and if there were a $k\in I$ such that $h_{\bar k}=\langle \psi|p_{\bar k}\bar p_k\rangle\not=0$ then, by introducing a normalized single qubit state $|\phi\rangle_k\propto h_I^*|p_k\rangle_k+h_{\bar k}^*|\bar p_k\rangle_k$, we would have $|\langle\psi |p_{\bar k}\phi_k\rangle|^2=|h_{\bar k}|^2+|h_I|^2>|h_I|^2$, which contradicts the definition of the closest product state as $|p_{\bar k}\phi_k\rangle$ is a product state.
The magic basis for a bipartite state coincides with its Schmidt decomposition. In general the magic basis for a given pure state is not unique and the one obtained from the closest product state only provides  us a possibility.

Under a magic basis for an entangled pure state $|\psi\rangle$ there is at least one $\alpha\subset I$ such that $h_\alpha\not=0$ and we introduce a nonnegative integer  \begin{equation}
m=\max_{\alpha\in\mathcal C}|\alpha|,\quad\mathcal C=\left\{\alpha\subset I| h_\alpha\not=0\right\}%\left\{|\alpha|\mid \alpha\subset I, h_\alpha\not=0\right\}
\end{equation}
for each pure state.  We refer to a subset   $A\in \mathcal C$ with $|A|=m$ as a {\it magic subset} for $|\psi\rangle$, which may not be unique. For a magic subset $A$ it holds  $h_A\not=0$ while  $h_B=0$ for any $B\subset I$ with $|A|<|B|<n$.  On the other hand, in a magic basis of $|\psi\rangle$ if the collection $\mathcal C$ is not empty then the state is entangled because local projection to $|0\rangle_k $ to each  qubit $k$ in a magic subset $A$ will leave those qubits in $\bar A$ in a GHZ-like state with nonzero coefficients $h_I$ and $h_A$, which is obviously entangled. By the definition of the magic basis we have $m\le n-2$.

\begin{table}
 \begin{equation*}\label{m1}
    \begin{array}{r@{\hskip .2cm}cc@{\hskip .2cm}cc@{\hskip .2cm}ccc}
    \hline\hline

      &a_{k0}& a_{k1}&b_{k0}& b_{k1}&\bar b_{k0}& \bar b_{k1}\\
       \hline
    k\in A &1&0&-\sin\gamma&\cos\gamma & \cos\gamma &\sin\gamma&\multirow{2}{*}{$\Big\}\ m=n-2$}\\
    k\in \bar A&1&0&q&r& -r^* &q\\

         \hline
    k\in A&1&0&c_k&z-1& 1-z^* &c^*_k&\multirow{3}{*}{$\Bigg\}\ m<n-2$}\\
   k\in S&1&y&1&yz&-yz^*&1\\
   k=v&e&1&-1&f&f^*&1\\

    \hline\hline
    \end{array}
    \end{equation*}
\caption{Two measurement settings $|a_k\rangle=a_{k0}|0\rangle_k+b_{k1}|1\rangle_k$ and $|b_k\rangle=b_{k0}|0\rangle_k+b_{k1}|1\rangle_k$ for each particle $k\in I$ for pure states in the Bell scenario $m=n-2$ (upper half) and the Hardy scenario (lower half) $m<n-2$. For each $k\in I$ state $|\bar b_{k}\rangle_k=\bar b_{k0}|0\rangle_k+\bar b_{k1}|1\rangle_k$ is orthogonal to $|b_k\rangle_k$.}
\end{table}

If $m=n-2$, i.e., there exists $A\subset I$ such that $h_A\not=0$ with $|A|=n-2$, we refer to this case as the Bell scenario.   In the upper part of Table.I two measurement settings $\{|a_k\rangle,|b_k\rangle\}$ are specified for each qubit $k\in I$ with  qubits in $A$ or $\bar A$ having the same pair of measurement settings, in which $q=\sqrt\lambda/(1+\lambda)$ and $r=ie^{-i\theta/2}\sqrt{1-q^2}$ with $\lambda=|h_A/h_I|>0$ and $h_A/h_I=\lambda e^{i\theta}$. We then have $
\langle H\rangle_\psi=|h_I|^2(1-2q^2-(n-2)\sin^2\gamma)-|\langle\psi |\bar b_I\rangle|^2$ where (see Appendix)
%$\langle \psi|a_I\rangle=h_I$, $\langle \psi|a_{\bar k} b_k\rangle=-h_I\sin\gamma $ for $k\in A$,  $\langle \psi|a_{\bar k} b_k\rangle=qh_I$ for $k\in \bar A$, and
\begin{eqnarray}
\langle\psi |\bar b_I\rangle=-\frac{h_Ie^{i\theta}}{1+\lambda}\cos^{m}\gamma
+\sum_{k=1}^{m}\sin^k\gamma \cos^{m-k}\gamma\hskip 1cm\cr
\times\mathop{\sum_{\beta\subseteq  A}}_{|\beta|=m-k}\left(r^{*2}h_{\beta\cup \bar A}-r^*q\sum_{{v}\in \bar A}h_{\beta\cup {v}}+q^2h_{\beta}\right).
\end{eqnarray}
If $\gamma=0$ we already have a violation $\langle H\rangle_\psi|_{\gamma=0}={|h_Ah_I|^2}/{(|h_A|+|h_I|)^2}>0$ and in this case two measurement directions for qubits in $A$ become identical. To have a nondegenerate pair of measurement settings we notice that  $
\langle H\rangle_\psi$ is a continuous function of $\gamma$ and there exists some small $\epsilon\not =0$ such that $\langle H\rangle_\psi|_{\gamma=\epsilon}>0$.

If $m<n-2$, i.e., there exists $A\subset I$ with $|A|=m$ such that $h_A\not=0$ while $h_B=0$ if $m<|B|<n$, we refer to this case as the Hardy scenario because the state exhibits Hardy-type nonlocality: the measurement settings can be so chosen that only the first term in Eq. (\ref{qd}) is nonzero. Consider the partition of the index set $I$ into 3 disjoint subsets $I=A\cup S\cup\{v\}$ with an arbitrary $v\in \bar A$ and $|S|=s=n-m-1\ge 2$. In the lower part of Table.I we have documented a family of measurement settings, with normalizations neglected,  determined by a real parameter $y\not=0$ and a complex parameter $z\not=1$ together with
$ f=h_I y^{-s}/{h_A}, e=-{h_A}y^{s}z/h_I,$
and for $k\in A$
\begin{equation}
\displaystyle c_k=\sum_{k^\prime\in S}\frac{h_{(A\setminus k)\cup k^\prime}}{yh_A}+\frac{h_{A\setminus k}}{h_A}-y^s\frac {h_{(A\setminus k)\cup v}}{h_I}z.
\end{equation}
Lengthy but direct calculations (see Appendix) yield $\langle\psi |a_{\bar k}b_k\rangle=0$ for all $k\in I$, $\langle \psi|a_I\rangle= y^sh_A(1-z)$, and
\begin{widetext}
\begin{eqnarray}
\langle \bar b_I|\psi\rangle&=&\Big(h_A^*+h_I^*(-yz)^sf\Big)(1-z)^m+\sum_{\alpha\subset A,\beta\subseteq S} (-yz)^{|\beta|}(1-z)^{|\alpha|}\left(h_{\alpha\cup\beta}^*+fh_{\alpha\cup\beta\cup v}^*\right)\prod_{k\in A-\alpha}c_{k}
\cr
&=&\frac{|h_I|^2}{h_{A}} (-z)^s(1-z)^m+\sum_{(\sigma,p,u,i,t)\in D}y^{t(s+1)+p+i-u-\sigma s}\ \Gamma_{puit}^{(\sigma)}(-z)^{t+p}(1-z)^{m-u}:=\sum_{k={-m-s}}^{(m+1)s}y^kL_k(z),
\end{eqnarray}
where $D=\{(\sigma,p,u,i,t)|0\le p\le s,p\le u\le m, 0\le i\le u,0\le t\le u-i,\sigma=0,1\}$ and we have denoted
\begin{equation}
\Gamma_{puit}^{(\sigma)}= \mathop{\sum_{\alpha \subseteq A}}_{|\alpha|=m-u}\mathop{\sum_{\beta\subseteq S}}_{|\beta|=p}G_{\alpha\beta}^{(\sigma)}\mathop{\sum_{\omega_1,\omega_2\subseteq A\setminus\alpha,\omega_1\cap\omega_2=\emptyset}}_{|\omega_1|=t,|\omega_2|=i}\left(\prod_{k\in A\setminus(\alpha\cup\omega_1\cup\omega_2)}\sum_{k^\prime\in S}\frac{h_{(A\setminus k)\cup k^\prime}}{h_A}\right)\left(\prod_{k\in \omega_1}\frac{h_{(A\setminus k)\cup v}}{h_I}\right)\left(\prod_{k\in \omega_2}\frac{h_{A\setminus k}}{h_A}\right)
\end{equation}
with $G_{\alpha\beta}^{(0)}=h_{\alpha\cup\beta}^*$ and $G_{\alpha\beta}^{(1)}={h_I}h_{\alpha\cup\beta\cup v}^*/h_A$. If we denote $D_0=\{(\sigma,p,u,i,t)\in D|u-i+\sigma s=t(s+1)+p\}$ then
\begin{equation}\label{f0}
L_0(z)=\frac{|h_I|^2}{h_{A}} (-z)^s(1-z)^m +\sum_{(\sigma,p,u,i,t)\in D_0}\Gamma^{(\sigma)}_{puit} (-z)^{t+p}(1-z)^{m-u}:=\sum_{k=0}^{n-1}l_k(-z)^k.
\end{equation}
\end{widetext}

We shall prove via {\it reductio ad absurdum} that there exists nonzero $y=y_0$ such that the algebraic equation $\langle \bar b_I|\psi\rangle=0$ of $z$ has one root $z=z_0\not=1$.  If  all the roots of $\langle \bar b_I|\psi\rangle=0$ were equal to 1 for any $y\not=0$, then $\langle \bar b_I|\psi\rangle$ as a polynomial of $z$ of degree $m+s=n-1$ would be proportional to $(1-z)^{n-1}$ and thus all the coefficients $L_k(z)$, especially
$L_0(z)$, would  be proportional to $(1-z)^{n-1}$ since $\{y^n\}_{n=-\infty}^{\infty}$ are linearly independent. On the one hand  we have $l_{n-1}=|h_I|^2/h_A$ and $l_{n-2}=ml_{n-1}$ for $L_0(z)$, taking into account the facts that $n-2>m$ and the sum term in Eq. (\ref{f0}) as a polynomial of $z$ is of degree at most $m$ because $t+p\le u$ in $D_0$ since $u\ge p$ for $t=0$ and $u-t-p=i+(t-\sigma)s\ge 0$ with $\sigma=0,1$ for $t\ge 1$. On the other hand for $(1-z)^{n-1}:=\sum_{k=0}^{n-1}l^\prime_k(-z)^k$ we have $l_{n-2}^\prime/l_{n-1}^\prime=n-1>m=l_{n-2}/l_{n-1}$, a contradiction.

Taking into account the normalization of $|a_I\rangle$ and parameters $y_0$ and $z_0$ determined above we can obtain the desired violation
\begin{equation}\label{v}
\langle H\rangle_\psi= \frac{\left|y_0^sh_Ah_I (1-z_0)\right|^2}{(1+y_0^2)^s\left(|h_I|^2+|y_0^sh_Az_0|^2\right)}>0.
\end{equation}
Two measurements directions in Table.I {may become identical for a qubit $k\in A$ if $c_k=0$, for all the qubits in $S$ if $z^*y^2=-1$, and for qubit $v$ if $f^*=e$, i.e., $-|h_I|^2=|h_A|^2y^2sz$. In these cases the degeneracy can be avoided by replacing $b_{k0}$ with $b_{k0}+x$, where $x$ is a real variable, while keeping $y_0$ and $z_0$ unchanged.} Since $\langle H\rangle_\psi$ depends on $x$ continuously and $\langle H\rangle_\psi|_{x=0}>0$, there exists small $\epsilon$ such that $\langle H\rangle_\psi|_{x=\epsilon}>0$ while two measurement directions are different for every qubit.

To sum up, for a given entangled pure $n$-qubit state $|\psi\rangle$ to violate Hardy's inequality we need only to find a magic basis and a magic subset $A$ for $|\psi\rangle$ and choose one set of the measurements defined in Table.I according to whether $|A|=m$ equals to $n-2$ or not. For an example, the $n$-qubit Dicke state $|S_k\rangle\propto\sum_{|\alpha|=k}|0_\alpha 1_{\bar\alpha}\rangle$ with $0<k<n$ belongs to the Bell scenario in the magic basis $\{|p_\alpha\bar p_{\bar\alpha}\rangle\}_{\alpha\subseteq I}$ determined by its closest product state  $|p\rangle^{\otimes n}$ with $|p\rangle\propto \sqrt k|0\rangle+\sqrt{n-k}|1\rangle$ and $|\bar p\rangle$ orthogonal to $|p\rangle$. The magic subset $A$ is any subset of $I$ with $n-2$ elements. Moreover, we have $$h_I^2=\binom nk k^k(n-k)^{n-k}/n^n$$ and $h_A=-h_I/(n-1)<0$ together with  $q=\sqrt{n-1}/n$ and $r=\sqrt{n^2-n+1}/n$, which lead to a violation $\langle H\rangle_{S_k}=h_I^2/n^2$ in the case of degenerate measurement settings $\gamma=0$.  The GHZ-like state $h_I^*|0_I\rangle+h_\emptyset^*|1_I\rangle$ with $h_Ih_\emptyset\not=0$, which is already expanded in a magic basis with the magic subset $A=\emptyset$ and $m=0$, belongs to the Hardy scenario. By taking $y_0=1$ the algebraic equation $\langle \bar b_I|\psi\rangle\propto |h_\emptyset|^2+|h_I|^2(-z)^{n-1}=0$ has a nonunital solution $z_0=-e^{i\pi/(n-1)}(|h_\emptyset|/|h_I|)^{2/(n-1)}$, which leads to a violation as given in Eq. (\ref{v}). As the last example the pure $n$-qubit state $|\psi\rangle\propto|0_I\rangle+|0_\alpha1_{\bar \alpha}\rangle+|1_I\rangle$ with $\alpha=\{1,2\}$ and $n=4j+1$ for $j\ge 1$ belongs to both the Bell and Hardy scenarios. First, the state is expressed already in a magic basis and the magic subset is $A=\alpha$ with $m=2<n-2$ since $n\ge 5$. By taking $v=\{n\}$ and $y_0=1$ we have $c_k=0$ for all $k\in A$ and $f=1, e=-z$. Since $z_0=i$ is a root of $\langle \bar b_I|\psi\rangle\propto (1-z)^2(1+(-z)^{n-3})=0$ we obtain a violation $\langle H\rangle_\psi=(3\times 2^{n-3})^{-1}$. Second, if $|0\rangle$ and $|1\rangle$ are exchanged for each qubit then we obtain  another magic basis with a magic subset $A=\bar \alpha$ with  $m=n-2$; i.e., the state $|\psi\rangle$ also belongs  to the Bell scenario with a violation $\langle H\rangle_\psi=1/12$ since $h_{\bar \alpha}=h_I=1/\sqrt3$.

{\it Violation for qudits.--- } Now we consider $n$ qudits, also labeled with $I$,  each of which may have a different number of energy levels. For a given pure $n$-qudit state $|\psi\rangle$  a magic basis can be defined similarly as in the case of qubits from its closest product state $|p_I\rangle=\otimes_{k\in I}|p_k\rangle_k$ satisfying $|\langle \psi|p\rangle|^2\le |\langle \psi|p_I\rangle|^2$ for any product state $|p\rangle$. We denote by $\mathcal C$ the collection of $\alpha\subset I$ such that for each $k\in\bar\alpha$ there exists a qudit state $|\bar p_k\rangle_k$ orthogonal to $|p_k\rangle_k$ such that
$\langle \psi|p_\alpha \bar p_{\bar\alpha}\rangle\not =0$. As long as  $|\psi\rangle$ is entangled the collection $\mathcal C$ is nonempty and vice versa and therefore the integer $m=\max_{\alpha\in \mathcal C}|\alpha|$ is well defined  such that
\begin{itemize}
\item[i.] There exists a magic subset $A\subset I$ with $|A|=m$ such that $h_A=\langle \psi|p_A \bar p_{\bar A}\rangle\not =0$  for some single qudit states $|\bar p_k\rangle_k$ orthogonal to $|p_k\rangle_k$ for each $k\in \bar A$ with $|\bar p\rangle_{\bar A}=\otimes_{k\in\bar A}|\bar p_k\rangle_k$;
\item[ii.]
For every subset $B\subset I$ with $m<|B|<n$ it holds $\langle \psi|p_B\phi_{\bar B}\rangle=0$ for all single qudit states $|\phi_k\rangle_k$ orthogonal to $|p_k\rangle_k$ for each $k\in \bar B$ with  $|p_B\rangle=\otimes_{k\in B}|p_k\rangle_k$ and $|\phi_{\bar B}\rangle=\otimes_{k\in \bar B} |\phi_k\rangle_k$.
\end{itemize}
Also we have $m\le n-2$  because if there were $k\in I$
such that $h_{\bar k}=\langle\psi|p_{\bar k}\phi_k\rangle\not=0$
for some qudit state $|\phi\rangle_k$ orthogonal to $|p_k\rangle_k$, then we would have $|\langle\psi|p_{\bar k}\phi^\prime_k \rangle|^2=|h_I|^2+|h_{\bar k}|^2>|h_I|^2$ with normalized state $|\phi^\prime\rangle_k\propto h_I^*|p_k\rangle_k+h^*_{\bar k}|\phi\rangle_k$, which contradicts the definition of the closest product state  as $|p_{\bar k}\phi^\prime_k \rangle$ is a product state.

For each qudit $k\in I$ we take $|p_k\rangle_k$ to be $|0\rangle_k$  and for each qubit $k\in \bar A$ we regard $|\bar p_k\rangle_k$, as it appeared in the definition of the magic subset $A$ (item i), to be $|1\rangle_k$ while for each qubit  $k\in A$ we take an arbitrary qudit state orthogonal to $|p_k\rangle_k$ to be $|1\rangle_k$. Thus we have picked out two orthogonal states for each qudit with the help of which we can {\it locally} project $n$ qudits to an $n$-qubit subspace.
Within this local $n$-qubit subspace we have effectively a set of $n$-qubits in a projected state (not normalized in general) in its magic basis with a magic subset $A$ satisfying $h_B=0$ as long as $|A|<|B|<n$. By choosing exactly the same measurement settings as specified in Table.I, we can obtain the desired violation to Hardy's inequality for an arbitrary entangled pure $n$-qudit state.

{\it Conclusions and discussions.--- }  We have proved Gisin's theorem in its most general form: every entangled pure state of a given number of particles, each of which may have a different number of energy levels,
violates one single Bell's inequality with two dichotomic observables for each observer.
Thus a strong equivalency between the quantum entanglement and Bell's nonlocality  is established for pure states.  In this sense Hardy's inequality is a more natural generalization of CHSH inequality to multiparticles. This is not surprising because Hardy's argument for nonlocality without inequality is de facto a state-dependent proof of quantum contextuality, which should manifest itself in any quantum state. It is argued in Ref.\cite{sc} that the postselection problem in the approach of Popescu and Rohrlich~\cite{pr} can be circumvented. In comparison we  need only a single Bell's inequality here. It is of interest to find the maximal violation of Hardy's inequality by a given pure state.

 We thank V. Scarani for bringing Ref. \cite{sc} to our attention  and stimulating discussions. This work is supported by National Research Foundation
and Ministry of Education, Singapore (Grant No.
WBS: R-710-000-008-271) and NSF of China (Grant No.
11075227).

\setcounter{equation}{0}
\renewcommand{\theequation}{S\arabic{equation}}
{\it Appendix.---} Here we shall present in details the calculations of the violations to Hardy's inequality by pure qubit states in the Bell scenario ($m=n-2$) and Hardy scenario ($m<n-2$), including especially the derivations of Eq.(5) and Eq.(7). We recall that we label $n$ qubits with the index set $I=\{1,,2,\ldots,n\}$ and for a subset $\alpha\subseteq I$ we denote $\bar\alpha=I\setminus \alpha$ and by $|\alpha|$ its number of elements. For a given pure $n$-qubit state $|\psi\rangle=\sum_{\alpha\subseteq I}h^*_\alpha|0_\alpha1_{\bar\alpha}\rangle$ with $h_\alpha=\langle\psi|0_\alpha1_{\bar\alpha}\rangle$ in its magic basis we have $h_I\not=0$ and  there exists  a magic subset $A\subset I$ such that $h_A\not =0$ and $h_B=0$ for all $B\subset I$ with $n>|B|>m=|A|$. 
\begin{widetext}
{\it Bell scenario: $I=A\cup \bar A$ with $|A|=m=n-2$. ---} According to the upper part of Table I we have
\begin{eqnarray}
|a_I\rangle&=&\bigotimes_{k\in I}|a_k\rangle_k=|0_I\rangle\\
|a_{\bar k}b_k\rangle&=&\bigotimes_{l\in \bar k}|a_l\rangle_l\otimes |b_k\rangle_k
=\left\{\begin{array}{ll}-\sin\gamma|0_I\rangle+\cos\gamma|0_{\bar k}1_k\rangle&\quad (k\in A)\cr\cr
q|0_I\rangle+r|0_{\bar k}1_k\rangle&\quad (k\in \bar A)\end{array}\right.\\
|\bar b_I\rangle&=&\bigotimes_{k\in I}|\bar b_k\rangle_k=\bigotimes_{k\in A}\Big(\cos\gamma|0\rangle_k+\sin\gamma|1\rangle_k\Big)\bigotimes_{l\in \bar A}\Big(-r^*|0\rangle_l+q|1\rangle_l\Big)\cr
&=&\sum_{\beta\subseteq A,\alpha\subseteq \bar A}(\cos\gamma)^{|\beta|}(\sin\gamma)^{m-|\beta|}(-r^*)^{|\alpha|}q^{2-|\alpha|}|0_{\beta\cup\alpha}1_{\overline{ \beta\cup\alpha}}\rangle
.
\end{eqnarray}
As a result we have $\langle\psi|a_I\rangle=h_I$ and $\langle\psi|a_{\bar k}b_k\rangle=-\sin\gamma h_I$ if $k\in A$ while $\langle\psi|a_{\bar k}b_k\rangle=q h_I$ if $k\in \bar A$, where we have used the fact that in the magic basis $h_{\bar k}=\langle \psi|0_{\bar k}1_k\rangle=0$ for arbitrary $k\in I$.
Taking into account the definitions $h_A=\lambda h_Ie^{i\theta}$, $q=\sqrt\lambda/(1+\lambda)$, and $r=ie^{-i\theta/2}\sqrt{1-q^2}$ in addition to the facts that $h_{\bar k}=0$ for all $k\in I$ and $\bar A$ has only two elements, we obtain Eq.(5)
\begin{eqnarray}
\langle\psi|\bar b_I\rangle&=&\Big(h_I(-r^*)^2+q^2 h_A\Big)\cos^m\gamma+\sum_{\beta\subset A,\alpha\subseteq \bar A}(\cos\gamma)^{|\beta|}(\sin\gamma)^{m-|\beta|}(-r^*)^{|\alpha|}q^{2-|\alpha|}h_{\beta\cup\alpha}\cr
&=&-\frac{h_Ie^{i\theta}}{1+\lambda}\cos^m\gamma+\sum_{k=1}^m(\cos\gamma)^{m-k}(\sin\gamma)^{k}\sum_{\beta\subset A,|\beta|=m-k}\left((-r^*)^2 h_{\beta\cup \bar A}-r^*q \sum_{v\in \bar A}h_{\beta\cup v}+q^2 h_{\beta}\right).
\end{eqnarray}

{\it Hardy scenario: $I=A\cup S\cup v$ with $|A|=m<n-2$ and $|S|=s=n-m-1$. ---} First of all we shall show that $\langle \psi|a_I\rangle= y^sh_A(1-z)$.  According to the lower part of Table~I we have
\begin{eqnarray}
|a_I\rangle&=&\bigotimes_{k\in I}| a_k\rangle_k=|0_A\rangle\bigotimes_{k\in S}\Big(|0\rangle_k+y|1\rangle_k\Big)\otimes\Big(e|0\rangle_v+|1\rangle_v\Big)=\sum_{\beta\subseteq S}y^{|\beta|}|0_{\overline{\beta\cup v}}1_{\beta\cup v}\rangle+e\sum_{\beta\subseteq S}y^{|\beta|}|0_{\bar\beta}1_{\beta}\rangle
.\end{eqnarray}
Because $A\subset\overline{\beta\cup v}\not=I$ for $\beta\subset S$ we have $n>|\overline{\beta\cup v}|>|A|=m$ and thus $h_{\overline{\beta\cup v}}=0$ if $\beta\not=S$ while $h_{\overline{S\cup v}}=h_A$. Because $A\subset \bar \beta\not=I$ for any nonempty $\beta\subseteq S$ we have $n>|\bar \beta|>|A|=m$ if $\beta$ is not empty and thus $h_{\bar \beta}=0$ if $\beta\not=\emptyset$ while $h_{\bar\emptyset}=h_I$. Considering $e=-zy^sh_A/h_I$ we obtain 
\begin{equation}
\langle \psi|a_I\rangle=y^s h_A+e h_I= y^sh_A(1-z).
\end{equation}
Secondly, we shall show that $\langle\psi |a_{\bar k}b_k\rangle=0$ for all $k\in I=A\cup S\cup v$. I) If $k=v$ then 
\begin{eqnarray}
|a_{\bar v}b_v\rangle&=&\bigotimes_{l\in \bar v}| a_l\rangle_l\otimes |b_v\rangle_v=|0_{A}\rangle\bigotimes_{l\in S}\Big(|0\rangle_l+y|1\rangle_l\Big)\otimes\Big(-|0\rangle_v+f|1\rangle_v\Big)=\sum_{\beta\subseteq S}y^{|\beta|}\Big(f|0_{\overline{\beta\cup v}}1_{\beta\cup v}\rangle
-|0_{\bar{\beta}}1_{\beta}\rangle\Big).
\end{eqnarray}
Since $h_{\overline{\beta\cup v}}=0$ for arbitrary $\beta\subset S$ and $h_{\bar \beta}=0$ for arbitrary nonempty $\beta\subseteq S$ we obtain ($f=h_Iy^{-s}/h_A$)
\begin{equation}
\langle \psi|a_{\bar v}b_v\rangle=y^s h_A f- h_I= 0.
\end{equation}
II) If $k\in A$ we have
\begin{eqnarray}
|a_{\bar k}b_k\rangle&=&\bigotimes_{l\in \bar k}| a_l\rangle_l\otimes |b_k\rangle_k=|0_{A\setminus k}\rangle\otimes\Big(c_k|0\rangle_k+(z-1)|1\rangle_k\Big)\bigotimes_{k\in S}\Big(|0\rangle_k+y|1\rangle_k\Big)\otimes\Big(e|0\rangle_v+|1\rangle_v\Big)\cr
&=&c_k|a_I\rangle+(z-1)\sum_{\beta\subseteq S}y^{|\beta|}\Big(|0_{\overline{\beta\cup\{k,v\}}}1_{\beta\cup\{k,v\}}\rangle
+e|0_{\overline{\beta\cup k}}1_{\beta\cup k}\rangle\Big)\end{eqnarray}
If $|\beta|<s$ we have $|\overline{\beta\cup k}|=n-|\beta|-1>n-s-1=m$  and thus $h_{\overline{\beta\cup k}}=0$. If $|\beta|<s-1$ we have $|\overline{\beta\cup\{k,v\}}|=n-|\beta|-2>m$ and thus $h_{\overline{\beta\cup\{k,v\}}}=0$. As a result
\begin{eqnarray}
\langle\psi|a_{\bar k}b_k\rangle&=&c_k\langle\psi|a_I\rangle+(z-1)\sum_{\beta\subseteq S}y^{|\beta|}\Big(h_{\overline{\beta\cup\{k,v\}}}+eh_{\overline{\beta\cup k}}\Big)\cr&=&c_ky^sh_A(1-z)+(z-1)y^s\Big(h_{\overline{S\cup\{k,v\}}}
+eh_{\overline{S\cup k}}\Big)+(z-1)y^{s-1}\sum_{\beta\subset S,|\beta|=s-1}h_{\overline{\beta\cup\{k,v\}}}\cr
&=&y^sh_A(1-z)\left(c_k-\frac{h_{A\setminus k}}{h_A}
-\frac{eh_{(A\setminus k)\cup v}}{h_A}-\sum_{k^\prime\in S}\frac{h_{(A\setminus k)\cup k^\prime}}{yh_A}\right)=0
\end{eqnarray}
where we have used the fact that $\beta\subset S$ with $|\beta|=s-1$ is equivalent to $\beta=S\setminus k^\prime$ with $k^\prime\in S$.
III) If $k\in S$ we have (from the lower part of Table I)
\begin{eqnarray}
|a_{\bar k}b_k\rangle&=&\bigotimes_{l\in \bar k}| a_l\rangle_l\otimes |b_k\rangle_k=|0_{A}\rangle\otimes\Big(|0\rangle_k+yz|1\rangle_k\Big)\bigotimes_{l\in S\setminus k}\Big(|0\rangle_l+y|1\rangle_l\Big)\otimes\Big(e|0\rangle_v+|1\rangle_v\Big)\cr
&=&|a_I\rangle+y(z-1)\sum_{\beta\subseteq S\setminus k}y^{|\beta|}\Big(e|0_{\overline{\beta\cup k}}1_{\beta\cup k}\rangle+|0_{\overline{\beta\cup\{k,v\}}}1_{\beta\cup\{k,v\}}\rangle
\Big).
\end{eqnarray}
Since $\beta\subseteq S\setminus k$ we have $|\beta|\le s-1$ and thus $h_{\overline{\beta\cup k}}=0$ for all $\beta\subseteq S\setminus k$ and $h_{\overline{\beta\cup\{k,v\}}}=0$ for all $\beta\subset S\setminus k$. As a result
\begin{equation}
\langle\psi|a_{\bar k}b_k\rangle=\langle\psi|a_I\rangle+y^s(z-1)h_A=0.
\end{equation}
Thirdly we shall derive Eq.(7). According to the lower part of Table I we have
\begin{eqnarray}
\langle\bar b_I|&=&\bigotimes_{k\in I}\langle \bar b_k|_k=\bigotimes_{k\in A}\Big(\langle 0|_k(1-z)+\langle1|_kc_k\Big)\bigotimes_{l\in S}\Big(-\langle 0|_lyz+\langle1|_l\Big)\otimes\Big(\langle 0|_vf+\langle1|_v\Big)\cr
&=&\sum_{\alpha\subseteq A,\beta\subseteq S}\Big(\langle 0_{{\alpha\cup\beta}}1_{\overline{\alpha\cup\beta}}|+\langle 0_{{\alpha\cup\beta\cup v}}1_{\overline{\alpha\cup\beta\cup v}}|f\Big)(-yz)^{|\beta|}(1-z)^{|\alpha|}\prod_{k\in A\setminus \alpha}c_k\end{eqnarray}
and therefore
\begin{eqnarray}
\langle\bar b_I|\psi\rangle&=&\sum_{\alpha\subseteq A,\beta\subseteq S}\Big(h^*_{{\alpha\cup\beta}}+fh^*_{{\alpha\cup\beta\cup v}}\Big)(-yz)^{|\beta|}(1-z)^{|\alpha|}\prod_{k\in A\setminus \alpha}c_k\cr
&=&\sum_{\beta\subseteq S}\Big(h^*_{{A\cup\beta}}+fh^*_{{A\cup\beta\cup v}}\Big)(-yz)^{|\beta|}(1-z)^{m}+\sum_{\alpha\subset A,\beta\subseteq S}\Big(h^*_{{\alpha\cup\beta}}+fh^*_{{\alpha\cup\beta\cup v}}\Big)(-yz)^{|\beta|}(1-z)^{|\alpha|}\prod_{k\in A\setminus \alpha}c_k\cr
&=&\Big(h^*_{{A}}+(-yz)^sfh^*_{I}\Big)(1-z)^{m}+\sum_{\alpha\subset A,\beta\subseteq S}\left(h^*_{{\alpha\cup\beta}}+\frac{h_Ih^*_{{\alpha\cup\beta\cup v}}}{y^sh_A}\right)(-yz)^{|\beta|}(1-z)^{|\alpha|}\prod_{k\in A\setminus \alpha}c_k\cr
&=&\left(h^*_{{A}}+(-z)^s\frac{|h_{I}|^2}{h_A}\right)(1-z)^{m}+\sum_{\alpha\subset A,\beta\subseteq S}\sum_{\sigma=0}^1G_{\alpha\beta}^{(\sigma)}y^{-\sigma s}(-yz)^{|\beta|}(1-z)^{|\alpha|}\prod_{k\in A\setminus \alpha}c_k\label{d}
\end{eqnarray}
where in the third equality we have used the fact that for $\beta\subseteq S$ $h_{A\cup \beta}\not=0$ if and only if $\beta$ is empty while $h_{A\cup \beta\cup v}\not=0$ if and only if $\beta=S$ and in the fourth equality we have introduced $G_{\alpha\beta}^{(0)}=h_{\alpha\cup\beta}^*$ and $G_{\alpha\beta}^{(1)}={h_I}h_{\alpha\cup\beta\cup v}^*/h_A$. To proceed we calculate
\begin{eqnarray}
\prod_{k\in A\setminus \alpha}c_k&=&\prod_{k\in A\setminus \alpha}\left(\sum_{k^\prime\in S}\frac{h_{(A\setminus k)\cup k^\prime}}{yh_A}+\frac{h_{A\setminus k}}{h_A}-y^s\frac {h_{(A\setminus k)\cup v}}{h_I}z\right)\cr &=&
\mathop{\sum_{\omega_1,\omega_2\subseteq A\setminus\alpha}}_{\omega_1\cap\omega_2=\emptyset}\left(\prod_{k\in (A\setminus\alpha)\setminus(\omega_1\cup\omega_2)}\sum_{k^\prime\in S}\frac{h_{(A\setminus k)\cup k^\prime}}{h_A}\right)\left(\prod_{k\in \omega_1}\frac{h_{(A\setminus k)\cup v}}{h_I}\right)\left(\prod_{k\in \omega_2}\frac{h_{A\setminus k}}{h_A}\right)\frac{(-y^sz)^{|\omega_1|}}{y^{|A\setminus\alpha|-|\omega_1\cup\omega_2|}}.\label{c}
\end{eqnarray}
Suppose $|\omega_1|=t$, $|\omega_2|=i$, $|\beta|=p$, and $|\alpha|=m-u$. Because $\beta\subseteq S$ we have $0\le p\le s$ and because $h_{\alpha\cup\beta}=0$ if $|\alpha\cup\beta|>m$ we have $p+m-u\le m$, i.e., $p\le u$. Furthermore we have $0\le i\le u$   and $0\le t\le u-i$ due to the facts that $\omega_2\subseteq A\setminus \alpha$, $\omega_1\cap\omega_2=\emptyset$, and $\omega_2\subseteq A\setminus \alpha$). Moreover since $\alpha\subset A$ we have $u\not=0$. We denote
$D=\{(\sigma,p,u,i,t)|0\le p\le s,p\le u\le m, 0\le i\le u,0\le t\le u-i,\sigma=0,1\}$. By substituting Eq.(\ref{c}) into Eq.(\ref{d}) we obtain
\begin{eqnarray}
&&\langle\bar b_I|\psi\rangle
-\left(h^*_{{A}}+(-z)^s\frac{|h_{I}|^2}{h_A}\right)(1-z)^{m}\cr&=&\sum_{(\sigma,p,u,i,t)\in D\cap\{u\not=0\}}y^{t(s+1)+p+i-u-\sigma s}(-z)^{t+p}(1-z)^{m-u}\ \mathop{\sum_{\alpha \subseteq A}}_{|\alpha|=m-u}\mathop{\sum_{\beta\subseteq S}}_{|\beta|=p}G_{\alpha\beta}^{(\sigma)}\cr &&\mathop{\sum_{\omega_1,\omega_2\subseteq A-\alpha,\omega_1\cap\omega_2=\emptyset}}_{|\omega_1|=t,|\omega_2|=i}\left(\prod_{k\in (A\setminus\alpha)\setminus(\omega_1\cup\omega_2)}\sum_{k^\prime\in S}\frac{h_{(A\setminus k)\cup k^\prime}}{h_A}\right)\left(\prod_{k\in \omega_1}\frac{h_{(A\setminus k)\cup v}}{h_I}\right)\left(\prod_{k\in \omega_2}\frac{h_{A\setminus k}}{h_A}\right)\cr
&:=&\sum_{(\sigma,p,u,i,t)\in D\cap\{u\not=0\}}y^{t(s+1)+p+i-u-\sigma s}(-z)^{t+p}(1-z)^{m-u}\ \Gamma_{puit}^{(\sigma)}.\end{eqnarray}
From the facts that $u=0$ leads to $p=t=i=0$ and $\Gamma_{0000}^{(0)}=h^*_A$, $\Gamma_{0000}^{(1)}=0$ Eq.(7) follows immediately. Finally we note that, since $t+i\le u\le m$, $p\le s$, and $t,p,i\ge 0$, we have bounds
\begin{equation}
-m-s\le -u-s\le t(s+1)+p+i-u-\sigma s\le ts+p\le (m+1)s
\end{equation}
with upper and lower bounds attained by $\{t=u=m,i=\sigma=0,p=s\}$ and \{$t=p=i=0,u=m,\sigma=1$\}, respectively.
\end{widetext}

\begin{thebibliography}{99}
\bibitem{bell} J.S. Bell, Physics  (N.Y.) {\bf 1}, 195 (1964).
\bibitem{werner} R.F. Werner Phys. Rev. A {\bf 40}, 4277 (1989).
\bibitem{gisin} N. Gisin, Phys. Lett. A {\bf 154}, 201 (1991).
\bibitem{CHSH} J.F. Clauser, M.A. Horne, A. Shimony, and R.A. Holt, Phys. Rev. Lett. {\bf 23}, 880 (1969).
\bibitem{pr} S. Popescu and D. Rohrlich, Phys. Lett. A {\bf 166}, 293 (1992).
\bibitem{zb}M. \.{Z}ukowski, C. Brukner, W. Laskowski, and M. Wiesniak, Phys. Rev. Lett. {\bf 88}, 210402 (2002).
\bibitem{chen} J.-L. Chen, C.F. Wu, L.C. Kwek, and C.H. Oh, Phys. Rev. Lett. {\bf 93}, 140407 (2004).
\bibitem{3} S.K. Choudhary, S. Ghosh, G. Kar, and R. Rahaman, Phys. Rev. A {\bf 81}, 042107 (2010).
\bibitem{gisinperes} N. Gisin and A. Peres, Phys. Lett. A {\bf 162}, 15 (1992).
\bibitem{chen2}J.-L. Chen, D.-L. Deng, and M.-G. Hu, Phys. Rev. A {\bf 77}, 060306(R) (2008).
\bibitem{shao} M. Li and S.-M. Fei, Phys. Rev. Lett. {\bf 104}, 240502 (2010).
\bibitem{commt}S.K. Choudhary, S. Ghosh, G. Kar, and R. Rahaman, Phys. Rev. Lett. {\bf 105}, 218901 (2010).
\bibitem{hardy} L. Hardy, Phys. Rev. Lett. {\bf 68}, 2981 (1992); L. Hardy, Phys. Rev. Lett. {\bf 71}, 1665 (1993).
\bibitem{mermin} N.D. Mermin, Am. J. Phys. {\bf 62}, 880 (1994). %Quantum mysteries refined
\bibitem{hardyn} J.L. Cereceda, Phys. Lett. A {\bf 327}, 433 (2004).% Hardy's nonlocality for generalized n-particle GHZ states
\bibitem{ks}S. Kochen and E.P. Specker, J. Math. Mech. {\bf 17}, 59 (1967).
%\bibitem{clifton} C. Pagonis and R. Clifton, Phys. Lett. A {\bf 168}, 100 (1992).
\bibitem{wu} X.-H. Wu and R.-H. Xie, Phys. Lett. A {\bf 211}, 129 (1996).
\bibitem{sym} Z. Wang and D. Markham, Phys. Rev. Lett. {\bf 108}, 210407 (2012).
\bibitem{gm} T.-C. Wei and P.M. Goldbart, Phys. Rev. A {\bf 68}, 042307 (2003).
\bibitem{sc} D. Cavalcanti, M.L. Almeida, V. Scarani, and A. Acin, Nature Communications {\bf 2}, 184 (2011).
\end{thebibliography}
\end{document}